# COMPLEX SUPPORT VECTOR MACHINE REGRESSION FOR ROBUST CHANNEL ESTIMATION IN LTE DOWNLINK SYSTEM

Anis Charrada[1] and Abdelaziz Samet[2]

[1] and [2] CSE Research Unit, Tunisia Polytechnic School, Carthage University,
Tunis, Tunisia .
[1]anis.charrada@gmail.com, [2]abdelaziz.samet@ept.rnu.tn

### ABSTRACT

*In this paper, the problem of channel estimation for LTE Downlink system in the environment of high mobility presenting non-Gaussian impulse noise interfering with reference signals is faced. The estimation of the frequency selective time varying multipath fading channel is performed by using a channel estimator based on a nonlinear complex Support Vector Machine Regression (SVR) which is applied to Long Term Evolution (LTE) downlink. The estimation algorithm makes use of the pilot signals to estimate the total frequency response of the highly selective fading multipath channel. Thus, the algorithm maps trained data into a high dimensional feature space and uses the structural risk minimization principle to carry out the regression estimation for the frequency response function of the fading channel. The obtained results show the effectiveness of the proposed method which has better performance than the conventional Least Squares (LS) and Decision Feedback methods to track the variations of the fading multipath channel.*

## 1. INTRODUCTION

Support Vector Machines (SVM) are a system for efficiently training the linear learning machines in the kernel-induced feature spaces while respecting the insights provided by the generalization theory and exploiting the optimization theory.

An important feature of these systems is that they produce sparse dual representations of the hypothesis while enforcing the learning biases suggested by the generalization theory which resulting in extremely efficient algorithms. This is due to the Karush–Kuhn–Tucker conditions which hold for the solution and play a fundamental role in the practical implementation and analysis of these machines. Another important feature of the Support Vector Machines technique is that the corresponding optimization problems are convex and hence have no local minima due to Mercer's conditions on the kernels. This fact and the reduced number of non-zero parameters mark a clear distinction between these systems and other pattern recognition algorithms such as neural networks.

Indeed, support vector machines have been proved to have superior performances in a large variety of digital communications and signal processing applications due to its improved generalization abilities and robustness against noise and interference. Based on the theory of SVM, Support Vector Regression (SVR) is a well established method for identification, estimation and prediction.

In this contribution, a proposed nonlinear SVR robust version for channel estimation specifically adapted to pilot-aided OFDM structure in the presence of impulse noise is presented and applied to LTE Downlink system. In fact, impulse noise can be present in a practical environment, so the channel becomes nonlinear with non-Gaussian impulse noise.

In this context, impulse noise can significantly influence the performance of the LTE Downlink system since the time of the arrival of an impulse is unpredictable and shapes of the impulses are not known and they vary considerably. Additionally, impulses usually have very high energy which can be much greater than the energy of the useful signal.

The principle of the proposed nonlinear complex SVR algorithm is to exploit the information provided by the pilot signals to estimate the channel frequency response. Thus, the proposed algorithm is developed in terms of the RBF (Radial Basis Function) kernel and applied to LTE downlink highly selective multipath fading channel. Simulation section illustrates the advantage of this algorithm over LS and Decision Feedback standard algorithms and compares the behaviour of these algorithms under high mobility conditions.

The scheme of the paper is as follows. We present a related work in section 2. Section 3 briefly introduces the LTE Downlink system model. Then, we present LS and Decision Feedback channel estimation methods with the formulation of the proposed nonlinear complex SVR method in section 4. The multipath mobile radio propagation channel model is described in Section 5. Section 6 shows the simulation results when comparing with LS and Decision Feedback algorithms. Finally, in section 7, conclusions are drawn.

## 2. RELATED WORK

The use of SVM has already been proposed to solve a variety of signal processing and digital communications problems, such that channel estimation by linear SVM in OFDM system which is presented in [1].

This study is specifically adapted to a pilots-based OFDM signal in the LTE Downlink context in a flat-fading channel and uses the block-type pilot structure. In this type, pilot tones are inserted into all subcarriers of pilot symbols with a period in time for channel estimation. This block-type pilot arrangement is suitable for slow fading channels. In fact, [1] consider a packet-based transmission, where each packet consists of a header at the beginning of the packet with a known training sequence or preamble to carry out channel estimation, followed by a certain number of OFDM data symbols. At the preamble, there are a number of OFDM symbols with a fixed number of pilot subcarriers in order to estimate the channels coefficient at pilot positions and then perform interpolation of the channel over all the OFDM symbols in the packet. However, in fast-fading channels, block-type pilot arrangement is not efficient and comb-type pilot arrangement is suitable especially in high mobility environments.

On the other hand, in [1], the channel's frequency response is estimated over a subset of pilot subcarriers and then interpolated over the remaining (data) subcarriers by using a DFT (Discret Fourier Transform) based technique with zero padding in the time domain. Therefore, the learning process can be more complex if the size of the pilot symbols is large, so the estimation task becomes slow.

The work of our paper focus on the nonlinear SVR applied to LTE Downlink system under high mobility conditions with comb-type pilot structure for fast fading channel. The LTE Downlink system under consideration requires an estimate of the frequency responses of data subchannels for each OFDM symbol. Therefore, the learning and estimation phases are repeated for each OFDM symbol in order to track the channel variation.

The use of the nonlinear SVR approach is needed for the channel estimation in deep fading environment. Indeed, Mercer's theorem announced that **x** in a finite dimension space (input space) can be mapped to a higher dimensional Hilbert space provided with a dot product through a nonlinear transformation $\boldsymbol{\varphi}(\cdot)$. Thus, a linear machine can be constructed in a higher dimensional space (feature space), but it stays nonlinear in the input space.

Note that most of the transformations $\boldsymbol{\varphi}(\cdot)$ are unknown, but the dot product of the corresponding spaces can be expressed as a function of the input vectors as

$$K(x_i, x_j) = <\boldsymbol{\varphi}(x_i), \boldsymbol{\varphi}(x_j)>. \tag{1}$$

These spaces are called Reproducing Kernel Hilbert Spaces (RKHS), and their dot products $K(x_i, x_j)$ are called Mercer kernels. Thus, it is possible to explicitly represent an SVM into a Hilbert spaces. The Mercer's theorem gives the condition that a kernel $K(x_i, x_j)$ must satisfy in order to be the dot product of a Hilbert space. The most popular kernel which satisfy Mercer's conditions is the RBF (Gaussian kernel) which is expressed as

$$K(x_i, x_j) = \exp - \frac{\|x_i - x_j\|^2}{2\sigma^2} \tag{2}$$

In this contribution, moreover, we use the indices of pilot positions (as input in learning phase) to estimate the channel frequency responses at these pilot positions (as out in learning phase), and then channel frequency responses at all subcarriers in each OFDM symbol can be obtained by SVR interpolation.

## 3. DOWNLINK LTE SYSTEM MODEL

Downlink LTE system model is based on OFDMA air interface transmission scheme. The basic idea of the OFDMA system is the use of the OFDM technique to divide the frequency spectrum into several orthogonal subcarriers. Those orthogonal frequency subcarriers are shared among all users using TDMA access technique.

Let us consider an LTE Downlink system which comprises $N$ subcarriers, occupying a bandwidth $B$. The corresponding OFDM system consists firstly of mapping binary data streams into complex symbols by means of QAM modulation. Then data are transmitted in frames by means of serial-to-parallel conversion. Some pilot symbols are inserted into each data frame which is modulated to subcarriers through IDFT. These pilot symbols are inserted for channel estimation purposes. The IDFT is used to transform the data sequence $X(k)$ into time domain signal as follow:

$$x(n) = IDFT_N\{X(k)\} = \sum_{k=0}^{N-1} X(k) e^{j\frac{2\pi}{N}kn}, \qquad n = 0, \cdots, N-1 \tag{3}$$

One guard interval is inserted between every two OFDM symbols in order to eliminate inter-symbol interference (ISI). This guard time includes the cyclically extended part of the OFDM symbol in order to preserve orthogonality and eliminate inter-carrier interference (ICI). It is well known that if the channel impulse response has a maximum of $L$ resolvable paths, then the GI must be at least equal to $L$ [2].

Thus, each OFDM symbol is transmitted in time $T$ and includes a cyclic prefix of duration $T_{cp}$. Therefore, the duration of each OFDM symbol is $T_u = T - T_{cp}$. Every two adjacent subcarriers are spaced by $\delta f = 1/T_u$. The output signal of the OFDM system is converted into serial signal by parallel to serial converter. A complex white Gaussian noise process $N(0, \sigma_w^2)$ with power spectral density $N_0/2$ is added through a frequency selective time varying multipath fading channel.

In a practical environment, impulse noise can be present, and then the channel becomes nonlinear with non Gaussian impulse noise. The impulse noise can significantly influence the performance of the OFDM communication system for many reasons. First, the time of the arrival of an impulse is unpredictable and shapes of the impulses are not known and they vary considerably. Moreover, impulses usually have very high amplitude, and thus high energy, which can be much greater than the energy of the useful signal [3].

The impulse noise is modeled as a Bernoulli-Gaussian process and it was generated with the Bernoulli-Gaussian process function $i(n) = v(n)\lambda(n)$ where $v(n)$ is a random process with Gaussian distribution and power $\sigma_{BG}^2$, and where $\lambda(n)$ is a random process with probability [4]

$$P_r(\lambda(n)) = \begin{cases} p & \lambda = 1 \\ 1-p, & \lambda = 0. \end{cases} \qquad (4)$$

At the receiver side, and after removing guard time, the discrete-time baseband OFDM signal for the system including impulse noise is

$$y(n) = \sum_{k=0}^{N-1} X(k)H(k)e^{j\frac{2\pi}{N}kn} + w(n) + i(n), \qquad n = 0,\cdots,N-1 \qquad (5)$$

where $y(n)$ are time domain samples and $H(k) = DFT_N\{h(n)\}$ is the channel's frequency response at the $k^{th}$ frequency. The sum of both terms of the AWGN noise and impulse noise constitute the total noise given by $z(n) = w(n) + i(n)$.

Let $\Omega_P$ the subset of $N_P$ pilot subcarriers. Over this subset, channel's frequency response can be estimated, and then interpolated over the remaining subcarriers $(N - N_P)$ which are interpolated by the nonlinear complex SVR algorithm. Thus, the OFDM system can be expressed as

$$y(n) = y^P(n) + y^D(n) + z(n)$$

$$= \sum_{k \in \{\Omega_P\}} X^P(k)H(k)e^{j\frac{2\pi}{N}kn} + \sum_{k \notin \{\Omega_P\}} X^D(k)H(k)e^{j\frac{2\pi}{N}kn} + z(n) \qquad (6)$$

where $X^P(k)$ and $X^D(k)$ are complex pilot and data symbol respectively, transmitted at the $k^{th}$ subcarrier. Note that, pilot insertion in the subcarriers of every OFDM symbol must satisfy the demand of the sampling theory and uniform distribution [5].

After DFT transformation, $y(n)$ becomes

$$Y(k) = DFT_N\{y(n)\} = \frac{1}{N}\sum_{n=0}^{N-1} y(n)\, e^{-j\frac{2\pi}{N}kn}, \quad k = 0,\cdots,N-1 \qquad (7)$$

Assuming that ISI are eliminated, therefore

$$Y(k) = X(k)H(k) + W(k) + I(k) = X(k)H(k) + e(k), \quad k = 0,\cdots,N-1 \qquad (8)$$

where $e(k)$ represents the sum of the AWGN noise $W(k)$ and impulse noise $I(k)$ in the frequency domain, respectively.

Equation (8) may be presented in matrix notation

$$\mathbf{Y} = \mathbf{XF}h + W + I = \mathbf{XH} + e \qquad (9)$$

where

$$X = diag(X(0), X(1), \cdots, X(N-1))$$
$$Y = [Y(0), \cdots, Y(N-1)]^T$$
$$W = [W(0), \cdots, W(N-1)]^T$$
$$I = [I(0), \cdots, I(N-1)]^T$$
$$H = [H(0), \cdots, H(N-1)]^T$$
$$e = [e(0), \cdots, e(N-1)]^T$$

$$F = \begin{bmatrix} W_N^{00} & \cdots & W_N^{0(N-1)} \\ \vdots & \ddots & \vdots \\ W_N^{(N-1)0} & \cdots & W_N^{(N-1)(N-1)} \end{bmatrix}$$

and
$$W_N^{i,k} = \left(\frac{1}{\sqrt{N}}\right) exp^{-j2\pi\left(\frac{ik}{N}\right)}. \tag{10}$$

As shown in Figure (1), each LTE radio frame duration is 10ms [6], which is divided into 10 subframes. Then, each subframe is further divided into two slots, each of 0.5 ms duration.

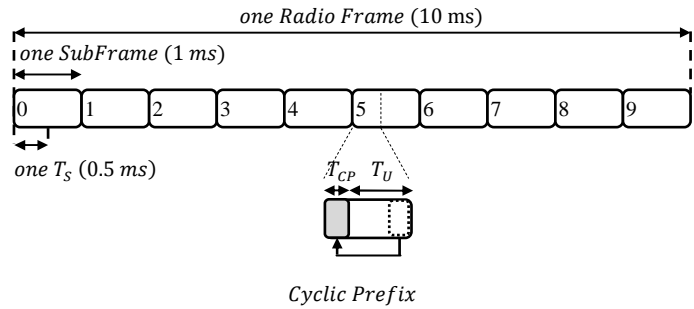

Figure 1. LTE Frame structure [6].

The physical resource block (PRB) consists of 12 subcarriers with frequency spacing of 15 kHz. In time domain, each PRP has one slot with either 6 or 7 OFDM symbols, depending on the chosen cyclic prefix, extended or normal. The transmission parameters of the LTE/OFDMA standard are shown in the following Table 1 [7].

Table 1. LTE OFDMA Parameters [7].

| Transmission BW (MHz) | 1.25 | 2.5 | 5 | 10 | 15 | 20 |
|---|---|---|---|---|---|---|
| Sub-frame duration (ms) | 0.5 | | | | | |
| Sub-carrier spacing (kHz) | 15 | | | | | |
| Sampling frequency | 1.92 | 3.84 | 7.68 | 15.36 | 23.04 | 30.72 |
| FFT size (*N*) | 128 | 256 | 512 | 1024 | 1536 | 2048 |
| Number of occupied sub-carriers | 76 | 151 | 301 | 601 | 901 | 1201 |

## 4. CHANNEL ESTIMATION

### 4.1. Least Squares Estimator

The principle of the channel least squares estimator is minimising the square distance between the received signal $Y$ and the original signal $X$ as follows

$$\min_{H^\dagger} J(H) = \min_{H^\dagger} \{|Y - XH|^2\}$$

where, $(.)^\dagger$ is the conjugate transpose operator.
By differentiating (11) with respect to $H^\dagger$ and finding the minima, we obtain the LS channel estimation which is given by [8]

$$\hat{H}_{LS} = X^{-1} Y \qquad (12)$$

### 4.2. Decision Feedback Estimator

OFDM channel estimation with decision feedback uses the reference symbols to estimate the channel response using LS algorithm. For each coming symbol $i$ and for each subcarrier $k$ for $k = 0, \cdots, N-1$, the estimated transmitted symbol is found from the previous $H_{k,i}$ according the formula

$$\hat{X}_{k,i+1} = \frac{Y_{k,i+1}}{\hat{H}_{k,i}} \qquad (13)$$

The estimated received symbols $\hat{X}_{k,i+1}$ are used to make the decision about the real transmitted symbol values $\tilde{X}_{k,i+1}$. The estimated channel response is updated by

$$\hat{H}_{k,i+1} = \frac{Y_{k,i+1}}{\tilde{X}_{k,i+1}} \qquad (14)$$

Therefore, $\hat{H}_{k,i+1}$ is used as a reference in the next symbol for the channel equalization.

### 4.3. Complex SVR Estimator

Let the OFDM frame contains $Ns$ OFDM symbols which every symbol includes $N$ subcarriers. The transmitting pilot symbols are $X^P = diag(X(s, m\,\Delta P)), m = 0,1,\cdots, N_P - 1$, where $s$ and $m$ are labels in time domain and frequency domain respectively, and $\Delta P$ is the pilot interval in frequency domain. Pilot insertion in the subcarriers of every OFDM symbol must satisfy the demand of sampling theory and uniform distribution [5].

The proposed channel estimation technique is based on nonlinear complex SVR algorithm which has two separate phases: learning phase and estimation phase. In learning phase, we estimate first the subchannels pilot symbols according to LS criterion to strike $min\ [(Y^P - X^P Fh)(Y^P - X^P Fh)^H]$ [9], as

$$\hat{H}^P = X^{P^{-1}} Y^P \qquad (15)$$

where $Y^P = Y(s, m\,\Delta P)$ and $\hat{H}^P = \hat{H}(s, m\,\Delta P)$ are the received pilot symbols and the estimated frequency responses for the $s^{th}$ OFDM symbol at pilot positions $m\,\Delta P$, respectively. Then, in the estimation phase and by the interpolation mechanism, frequency responses of data subchannels can be determined. Therefore, frequency responses of all the OFDM subcarriers are

$$\hat{H}(s, q) = f\left(\hat{H}^P(s, m\,\Delta P)\right) \qquad (16)$$

where $q = 0, \cdots, N - 1$, and $f(\cdot)$ is the interpolating function, which is determined by the nonlinear complex SVR approach.

Linear approaches cannot achieve high estimation precision in high mobility environments where the fading channels present very complicated nonlinearities. Therefore, we adapt here a nonlinear complex SVR method since SVM is superior in solving nonlinear, small samples and high dimensional pattern recognition [5]. Thus, we map the input vectors to a higher dimensional feature space $\mathcal{H}$ (possibly infinity) by means of nonlinear transformation $\boldsymbol{\varphi}(\cdot)$. So, the regularization term is referred to the regression vector in the RKHS. The following regression function is then

$$\widehat{H}(m\,\Delta P) = \boldsymbol{w}^T \boldsymbol{\varphi}(m\,\Delta P) + b + e_m, \quad m = 0, \cdots, N_P - 1 \quad (17)$$

where $\boldsymbol{w}$ is the weight vector, $b$ is the bias term well and residuals $\{e_m\}$ account for the effect of both approximation errors and noise. In the SVM framework, the optimality criterion is a regularized and constrained version of the regularized Least Squares criterion. In general, SVM algorithms minimize a regularized cost function of the residuals, usually the Vapnik's $\varepsilon-insensitivity$ cost function [4].

To improve the performance of the estimation algorithm, a robust cost function is introduced which is $\varepsilon$ -Huber robust cost function [10], given by

$$\mathcal{L}^\varepsilon(e_m) = \begin{cases} 0, & |e_m| \leq \varepsilon \\ \frac{1}{2\gamma}(|e_m| - \varepsilon)^2, & \varepsilon \leq |e_m| \leq e_C \\ C(|e_m| - \varepsilon) - \frac{1}{2}\gamma C^2, & e_C \leq |e_m| \end{cases} \quad (18)$$

where $e_C = \varepsilon + \gamma C$, $\varepsilon$ is the insensitive parameter which is positive scalar that represents the insensitivity to a low noise level, parameters $\gamma$ and $C$ control essentially the trade-off between the regularization and the losses, and represent the relevance of the residuals that are in the linear or in the quadratic cost zone, respectively. The cost function is linear for errors above $e_C$, and quadratic for errors between $\varepsilon$ and $e_C$. Note that, errors lower than $\varepsilon$ are ignored in the $\varepsilon - insensitivite$ zone. The quadratic cost zone uses the $L_2 - norm$ of errors, which is appropriate for Gaussian noise, and the linear cost zone limits the effect of sub-Gaussian noise [1]. Therefore, the $\varepsilon$ -Huber robust cost function can be adapted to different types of noise.

Let $\mathcal{L}^\varepsilon(e_m) = \mathcal{L}^\varepsilon(\mathcal{R}(e_m)) + \mathcal{L}^\varepsilon(\mathfrak{I}(e_m))$ since $\{e_m\}$ are complex, where $\mathcal{R}(\cdot)$ and $\mathfrak{I}(\cdot)$ represent real and imaginary parts, respectively. Now, we can state the primal problem as

$$min\ \frac{1}{2} \|\boldsymbol{w}\|^2 + \frac{1}{2\gamma} \sum_{m \in I_1} (\xi_m + \xi_m^*)^2 + C \sum_{m \in I_2} (\xi_m + \xi_m^*) + \frac{1}{2\gamma} \sum_{m \in I_3} (\zeta_m + \zeta_m^*)^2$$
$$+ C \sum_{m \in I_4} (\zeta_m + \zeta_m^*) - \frac{1}{2} \sum_{m \in I_2, I_4} \gamma C^2 \quad (19)$$

constrained to

$$\mathcal{R}\big(\widehat{H}(m\,\Delta P) - \boldsymbol{w}^T \boldsymbol{\varphi}(m\,\Delta P) - b\big) \leq \varepsilon + \xi_m$$

$$\mathfrak{I}\big(\widehat{H}(m\,\Delta P) - \boldsymbol{w}^T \boldsymbol{\varphi}(m\,\Delta P) - b\big) \leq \varepsilon + \zeta_m$$

$$\mathcal{R}(-\widehat{H}(m\,\Delta P) + \mathbf{w}^T\boldsymbol{\varphi}(m\,\Delta P) + b) \leq \varepsilon + \xi_m^*$$

$$\mathfrak{I}(-\widehat{H}(m\,\Delta P) + \mathbf{w}^T\boldsymbol{\varphi}(m\,\Delta P) + b) \leq \varepsilon + \zeta_m^*$$

$$\xi_m^{(*)}, \zeta_m^{(*)} \geq 0 \qquad (20)$$

for $m = 0, \cdots, N_P - 1$, where $\xi_m$ and $\xi_m^*$ are slack variables which stand for positive and negative errors in the real part, respectively. $\zeta_m$ and $\zeta_m^*$ are the errors for the imaginary parts. $I_1, I_2, I_3$ and $I_4$ are the set of samples for which:

$I_1$ : real part of the residuals are in the quadratic zone;

$I_2$ : real part of the residuals are in the linear zone;

$I_3$ : imaginary part of the residuals are in the quadratic zone;

$I_4$ : imaginary part of the residuals are in the linear zone.

To transform the minimization of the primal functional (19) subject to constraints in (20), into the optimization of the dual functional, we must first introduce the constraints into the primal functional.

Then, by making zero the primal-dual functional gradient with respect to $\varpi_i$, we obtain an optimal solution for the weights

$$\mathbf{w} = \sum_{m=0}^{N_P-1} \psi_m\,\boldsymbol{\varphi}(m\,\Delta P) = \sum_{m=0}^{N_P-1} \psi_m\,\boldsymbol{\varphi}(P_m) \qquad (21)$$

where $\psi_m = (\alpha_{\mathcal{R},m} - \alpha_{\mathcal{R},m}^*) + j(\alpha_{I,m} - \alpha_{I,m}^*)$ with $\alpha_{\mathcal{R},m}, \alpha_{\mathcal{R},m}^*, \alpha_{I,m}, \alpha_{I,m}^*$ are the Lagrange multipliers (or dual variables) for real and imaginary part of the residuals and $P_m = (m\,\Delta P)$, $m = 0, \cdots, N_P - 1$ are the pilot positions.

Let the Gram matrix defined by

$$\mathbf{G}(u,v) = <\boldsymbol{\varphi}(P_u), \boldsymbol{\varphi}(P_v)> = K(P_u, P_v) \qquad (22)$$

where $K(P_u, P_v)$ is a Mercer's kernel which represent the RBF kernel matrix which allows obviating the explicit knowledge of the nonlinear mapping $\boldsymbol{\varphi}(\cdot)$. A compact form of the functional problem can be stated in matrix format by placing optimal solution $\mathbf{w}$ into the primal dual functional and grouping terms. Then, the dual problem consists of

$$max \; -\frac{1}{2}\boldsymbol{\psi}^H(\mathbf{G} + \gamma\mathbf{I})\boldsymbol{\psi} + \mathcal{R}(\boldsymbol{\psi}^H Y^P) - (\boldsymbol{\alpha}_{\mathcal{R}} + \boldsymbol{\alpha}_{\mathcal{R}}^* + \boldsymbol{\alpha}_I + \boldsymbol{\alpha}_I^*)\mathbf{1}\varepsilon \qquad (23)$$

constrained to

$$0 \leq \alpha_{\mathcal{R},m}, \alpha_{\mathcal{R},m}^*, \alpha_{I,m}, \alpha_{I,m}^* \leq C \qquad (24)$$

where $\boldsymbol{\psi} = [\psi_0, \cdots, \psi_{N_P-1}]^T$ ; $\mathbf{I}$ and $\mathbf{1}$ are the identity matrix and the all-ones column vector, respectively; $\boldsymbol{\alpha}_{\mathcal{R}}$ is the vector which contains the corresponding dual variables, with the other subsets being similarly represented. The weight vector can be obtained by optimizing (23) with respect to $\alpha_{\mathcal{R},m}, \alpha_{\mathcal{R},m}^*, \alpha_{I,m}, \alpha_{I,m}^*$ and then substituting into (21).

Therefore, and after learning phase, frequency responses at all subcarriers in each OFDM symbol can be obtained by SVR interpolation

$$\hat{H}(k) = \sum_{m=0}^{N_P-1} \psi_m K(P_m, k) + b \quad (25)$$

for $k = 1, \cdots, N$. Note that, the obtained subset of dual multipliers which are nonzero will provide with a sparse solution. As usual in the SVM framework, the free parameter of the kernel and the free parameters of the cost function have to be fixed by some a priori knowledge of the problem, or by using some validation set of observations [4].

## 5. MULTIPATH CHANNEL MODEL

We consider the channel impulse response of the frequency-selective fading multipath channel model which can be written as

$$h(\tau, t) = \sum_{l=0}^{L-1} h_l(t) \delta(t - \tau_l) \quad (26)$$

where $h_l(t)$ is the impulse response representing the complex attenuation of the $l^{th}$ path, $\tau_l$ is the random delay of the $l^{th}$ path and $L$ is the number of multipath replicas. The specification parameters of an extended vehicular A model (EVA) for downlink LTE system with the excess tap delay and the relative power for each path of the channel are shown in table 2. These parameters are defined by 3GPP standard [11].

Table 2. Extended Vehicular A model (EVA) [11].

| Excess tap delay [ns] | Relative power [dB] |
|---|---|
| 0 | 0.0 |
| 30 | -1.5 |
| 150 | -1.4 |
| 310 | -3.6 |
| 370 | -0.6 |
| 710 | -9.1 |
| 1090 | -7.0 |
| 1730 | -12.0 |
| 2510 | -16.9 |

## 6. SIMULATION RESULTS

In order to demonstrate the effectiveness of our proposed method and evaluate the performance, two objective criteria, the signal-to-noise ratio (SNR) and signal-to-impulse ratio (SIR) are used. The SNR and SIR are given by [4]

$$SNR_{dB} = 10 log_{10} \left( \frac{E\{|y(n) - w(n) - i(n)|^2\}}{\sigma_w^2} \right) \quad (27)$$

and

$$SIR_{dB} = 10 log_{10} \left( \frac{E\{|y(n) - w(n) - i(n)|^2\}}{\sigma_{BG}^2} \right) \quad (28)$$

Then, we simulate the OFDM downlink LTE system with parameters presented in table 3. The nonlinear complex SVR estimate a number of OFDM symbols in the range of 140 symbols, corresponding to one radio frame LTE.

For the purpose of evaluation the performance of the nonlinear complex SVR algorithm under high mobility conditions, we consider the scenario for LTE Downlink system for a mobile speed at 350 Km/h. Figure (2) presents the variations in time and in frequency of the channel frequency response for the considered scenario.

Table 3. Parameters of simulations [6], [7] and [12].

| Parameters | Specifications |
|---|---|
| OFDM system | LTE/Downlink |
| Constellation | 16-QAM |
| Mobile Speed (Km/h) | 350 |
| $T_s$ (μs) | 72 |
| $f_c$ (GHz) | 2.15 |
| $\delta f$ (KHz) | 15 |
| $B$ (MHz) | 5 |
| Size of DFT/IDFT | 512 |
| Number of paths | 9 |

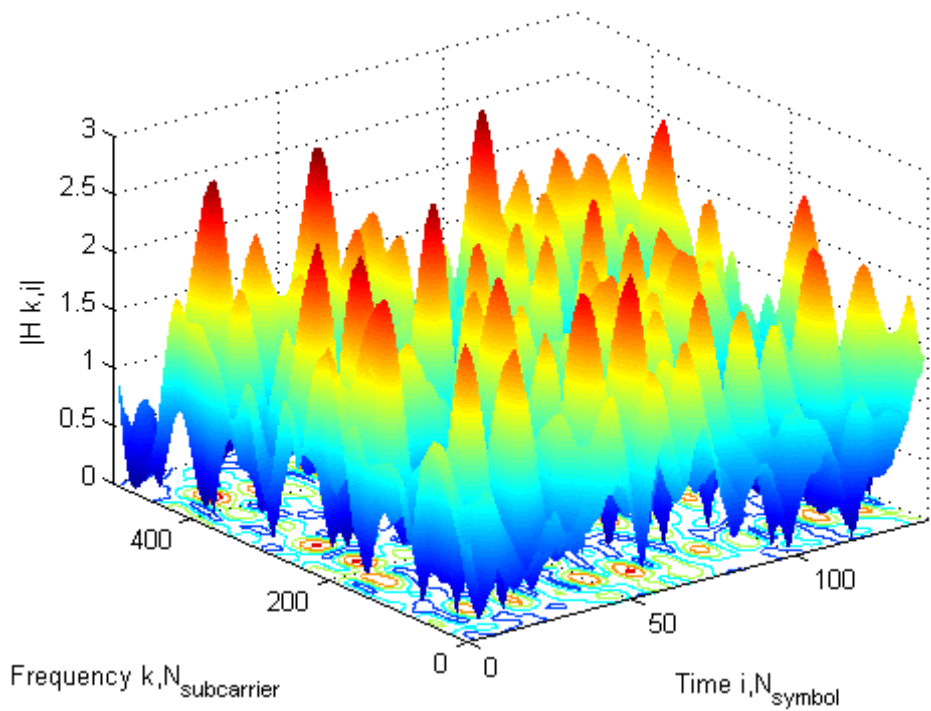

Figure 2. Variations in time and in frequency of the channel frequency response for mobile speed at 350 Km/h.

Figure (3) presents the performance of the LS, Decision Feedback and nonlinear complex SVR algorithms for the scenario without non-Gaussian impulse noise. A poor performance is noticeably exhibited by LS and Decision Feedback while better performance is observed with nonlinear complex SVR.

International Journal of Computer Networks & Communications (IJCNC) Vol.4, No.1, January 2012

Figure (4) shows the performance of LS, Decision Feedback and nonlinear complex SVR techniques in the presence of non-Gaussian impulse noise with p=.1 for SNR = 20 dB as a function of SIR which is ranged from -15 to 15 dB. This figure confirms that nonlinear complex SVR algorithm performs better than LS and Decision Feedback algorithms in the environment presenting impulse noise with high mobility conditions.

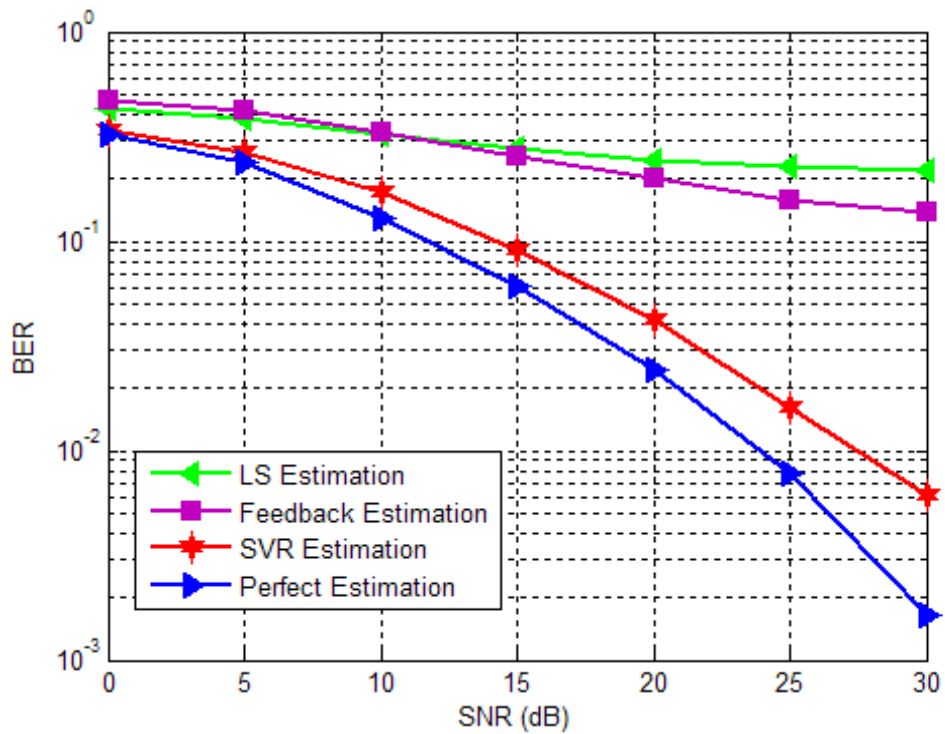

Figure 3. BER as a function of SNR for a mobile speed at 350 Km/h without impulse noise.

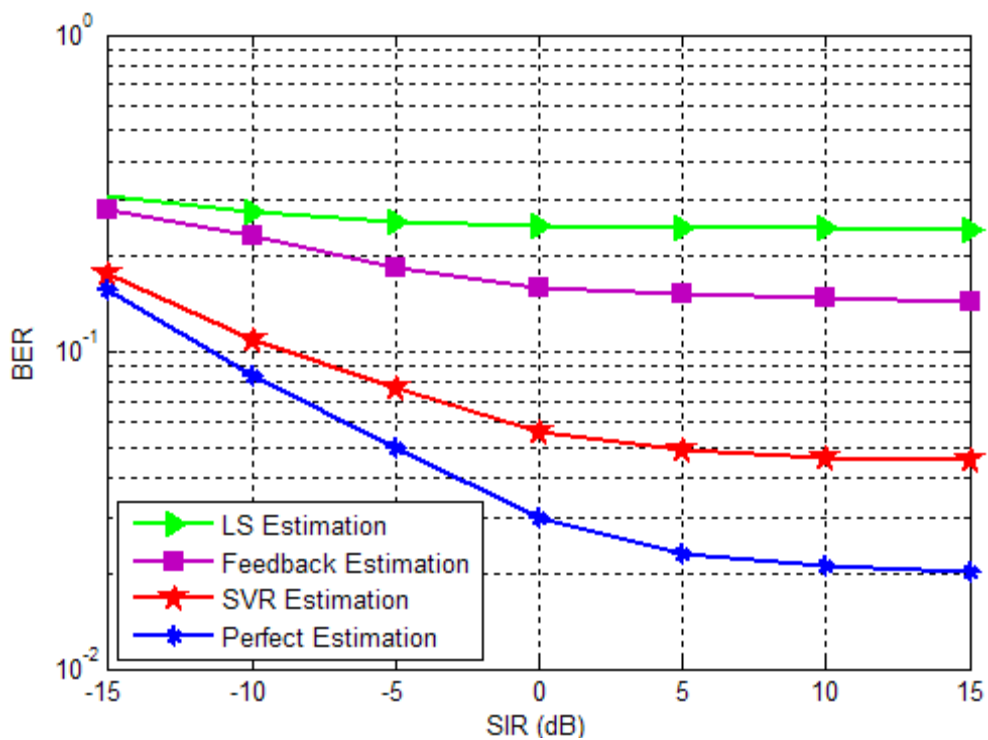

Figure 4. BER as a function of SIR for a mobile speed at 350 Km/h for SNR=20 dB with p=.1.

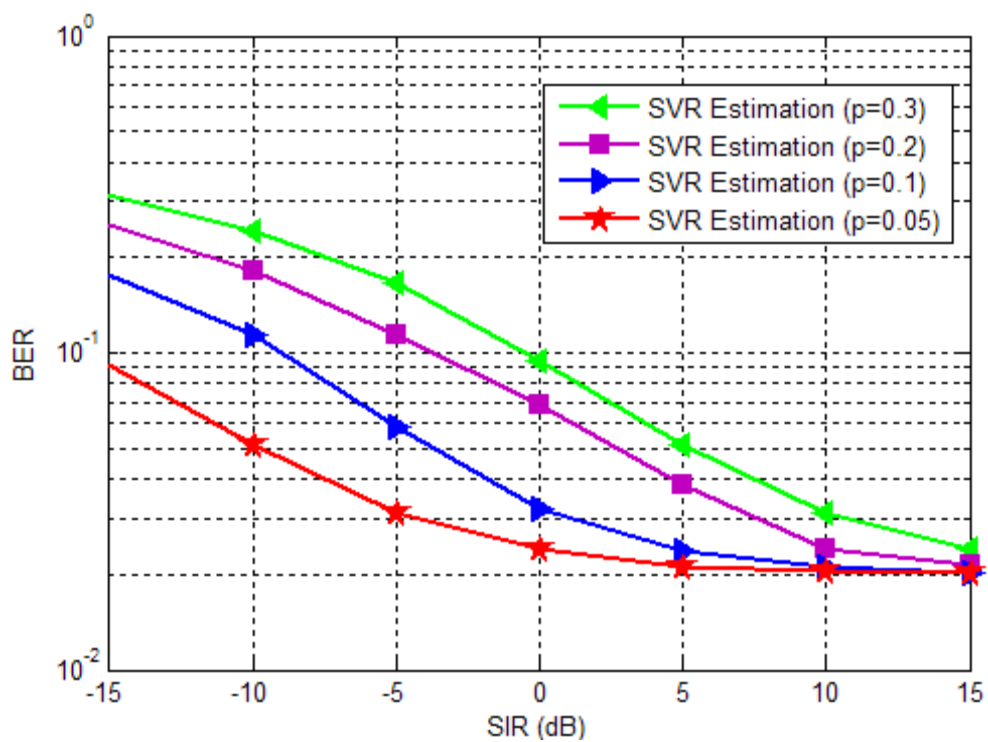

Figure 5.  BER as a function of SIR for a mobile speed at 350 Km/h for SNR=30 dB

with different values of p.

Accordingly, we take into account the impulse noise with different values of $p$ which was added to the reference signals with various rates of SIR. Figure (5) compares the behavior of the nonlinear complex SVR with different values of $p$ of the impulse noise for SNR = 30 dB as a function of SIR for the scenario under consideration. The comparison of these curves reveals that nonlinear complex SVR with $p = .05$ outperforms other values of $p$ especially in low SIR which confirms the capabilities of the proposed method in the presence of nonlinear noise.

## 7. CONCLUSION

This paper describes a nonlinear complex SVR based channel estimation technique for a downlink LTE system in the presence of Gaussian noise and impulse noise interfering with OFDM pilot symbols in high mobility environment. The proposed method is based on learning process that uses training sequence to estimate the channel variations. Our formulation is based on nonlinear complex SVR specifically developed for reference-based OFDMA systems. Simulations have confirmed the capabilities of the proposed nonlinear complex OFDM-SVR estimator in the presence of Gaussian and non-Gaussian impulse noise interfering with the reference symbols for a high mobile speed when compared to conventional LS and Decision feedback methods for a highly selective time varying multipath fading channel.